\documentclass[twocolumn,aps]{revtex4}
\newcommand{\be}{\begin{equation}}
\newcommand{\ee}{\end{equation}}
\begin{document}
\title{Comment on ``Strength and genericity of singularities
in Tolman-Bondi-de Sitter collapse''  
and a note on central singularities}
\author{
{Brien C. Nolan$^{(1)}$, Filipe C. Mena$^{(2)}$ and
S\'{e}rgio M. C. V. Gon\c{c}alves$^{(3)}$}
\\ {\em $^{(1)}$ School of Mathematical Sciences,
Dublin City University, Glasnevin, Dublin 9, Ireland}  \\
{\em $^{(2)}$School of Mathematical Sciences, Queen
Mary, University of London, Mile End Rd, London E1 4NS, U.K.}\\
{\em $^{(3)}$Theoretical Astrophysics, California
Institute of Technology, Pasadena, CA 91125, U.S.A.}}
\date{16th August 2001}
\begin{abstract}
It has been claimed in \cite{sergio1} that the Lemaitre--Tolman--Bondi--de
Sitter solution always admits future-pointing radial time-like
geodesics emerging from the shell-focussing singularity, regardless
of the nature of the (regular) initial data. This is despite the
fact that some data rule out the emergence of future pointing
radial null geodesics. We correct this claim and show that in
general in spherical symmetry, the absence of radial null
geodesics emerging from a central singularity is sufficient to
prove that the singularity is censored.
\end{abstract}
\maketitle

The central or shell-focussing singularity which occurs in the
gravitational collapse of spherical dust in the presence of a
positive cosmological constant has been studied in
\cite{sergio1,sergio2}. In the latter paper, it was shown that the
existence or otherwise of radial null geodesics emerging from the
singularity depends on the initial data in much the same way as
this dependence occurs in the asymptotically flat case. However,
in \cite{sergio1}, it is claimed that there are future pointing
radial time-like geodesics emerging from the singularity for {\em
all} regular initial data. This is in contrast to the
asymptotically flat case and is somewhat surprising, given that
radial null geodesics are the `fastest' causal geodesics
available~\cite{reason}, and so have the best chance of emerging
from a singularity (this statement is made more rigorous below).
We show here that the analysis of \cite{sergio1} is incomplete,
and thus one of the results asserted---that the central
singularity is always visible along radial time-like geodesics,
regardless of the initial data---has not been proven. We then give
a counter-example and a general result, which show that the
assertion is incorrect.

The space-time in question is the marginally bound
Lemaitre-Tolman-Bondi with line element given by
\cite{Lemaitre} \be ds^2=-dt^2+{R^\prime}^2dr^2+R^2d\Omega^2,\ee
where $'\equiv\partial_r$, and $d\Omega^2$ is the canonical metric
of the unit two-sphere. The field equations for dust in the
presence of a positive cosmological constant $\Lambda$ yield
\begin{eqnarray}
R^3(t,r)=\frac{6m}{\Lambda}\sinh^2T(t,r), \label{ard} \\
T(t,r)=\frac{\sqrt{3\Lambda}}{2}[t_c(r)-t],
\end{eqnarray}
where $m=m(r)$ and $t_c(r)$ are functions determined by the
initial data, which are imposed at time $t=0$. The dust sphere
collapses to a singularity with zero radius ($R=0)$ at time
$t=t_c(r)$; the singularity at $r=0$ is called the shell-focussing
singularity; we will simply refer to it as `the singularity'. We
can exploit coordinate freedom to set $R(0,r)=r$ and so obtain
\[ t_c(r)=\frac{2}{\sqrt{3\Lambda}}\sinh^{-1}\left(\sqrt{\frac{\Lambda
r^3}{6m}}\right).\] The following derivatives are of relevance for the
analysis that follows:
\begin{eqnarray}
R^\prime&=&
R\left(\frac{m^\prime}{3m}+\sqrt{\frac{\Lambda}{3}}t_c^\prime\coth T\right),\\
R^{\prime\prime}&=&\frac{(R^\prime)^2}{R}+R\left(\frac{m^{\prime\prime}}{3m}
-\frac13\left(\frac{m^\prime}{m}\right)^2\right)\nonumber\\
&& +R\sqrt{\frac{\Lambda}{3}}\left(t_c^{\prime\prime}\coth
T-\frac{\sqrt{3\Lambda}}{2}\frac{(t_c^\prime)^2}{\sinh^2T}\right),\\
{\dot R}^\prime&=&-\sqrt{\frac{\Lambda}{3}}\coth TR^\prime
+\frac{\Lambda}{2}R\frac{t_c^\prime}{\sinh^2T}.
\end{eqnarray}
The overdot denotes differentiation with respect to $t$.
The equations governing a radial time-like geodesic with tangent
$K^a=\frac{dx^a}{d\tau}$ (where $\tau$ is proper time) may be written as
\begin{eqnarray}
K^t=\pm\sqrt{1+(R^\prime)^2(K^r)^2},\\
{\dot K^r}R^\prime+2K^r{\dot R}^\prime+\frac{K^r}{K^t}(K^r)^\prime
R^\prime+\frac{(K^r)^2}{K^t}R^{\prime\prime}=0.
\end{eqnarray}
In \cite{sergio1}, a proof of the existence of a solution of these
equations emerging from the singularity is attempted by assuming
the ansatze
\begin{eqnarray}
t_{RTG}(r)=t_0+br^p,\label{a1} \\
R^\prime=a_1r^q,\\
{\dot R}^\prime=a_2r^{q-p},\\
R^{\prime\prime}=a_3r^{q-1},\\
K^r\propto(t-t_0)^\alpha r^\beta,
\end{eqnarray}
where $t_0=t_c(0)$ is the time of the shell-focussing singularity
and $b,p,q$ are positive constants. It is then claimed that these
are consistent with the geodesic equations and so indicate the
existence of a solution representing a radial time-like geodesic
(RTG) emerging from the singularity. Furthermore, it is claimed
that this result follows independently of the initial data $m(r)$.

A vital part of this consistency check results in $p=1+q$. To see
that this condition may fail, we consider the example \be m(r) =
m_0r^3 + m_1r^7.\ee The lower power here is required for
regularity of the initial data, and the higher power ensures that
that there are no radial null geodesics emerging from the
singularity \cite{sergio2}. This choice is included in the class
of mass functions $m(r)$ considered in \cite{sergio1}. Along an
RTG emerging from the singularity, we must have $T\geq0$ for $r$
sufficiently small, with equality only at $r=0$. Then examining
the leading order behaviour in
\[ T(r)=\frac{\sqrt{3\Lambda}}{2}(t_c(r)-t_0-br^p),\]
for the mass function given, we deduce that $p\geq4$ and that
\[ T(r)\sim T_0r^4,\]
for some positive $T_0$ (the fact that $T\geq0$ is vital here).
The functional dependence here and below indicates evaluation
along the geodesic. This asymptotic behaviour can be fed into the
expressions above for $R$ and its derivatives and yields
\begin{eqnarray}
R(r)\sim R_0r^{11/3},\\
R^\prime(r)\sim R_1r^{8/3},\\
{\dot R}^\prime(r)\sim R_2r^{-4/3},\\
R^{\prime\prime}(r)\sim R_3 r^{5/3}.
\end{eqnarray}
Comparing with the ansatze above, we see that $q=8/3$ and $p=4$.
However this violates the consistency condition $p=q+1$,
indicating that such a solution cannot in fact exist. We note that
two assumptions made here played a vital role: (i) the mass
function $m(r)$ excludes radial null geodesics emerging from the
singularity, and (ii) the RTG emerges into the regular region of
space-time $T>0$.

The crucial point that is missing in~\cite{sergio1}, is that the
parameters $b$ and $p$ [cf. Eq. (28) in~\cite{sergio1}] are {\em
not} independent of the initial data, as implicitly assumed
therein. In fact, we must have $p\geq n$ [where $n$ signals the
first-non-vanishing derivative of the initial central density
distribution, $\rho_{n}\equiv(\partial^{n}\rho/\partial
r^{n})_{r=0}$], or else the geodesic thus constructed will not
belong in the spacetime. When this inequality saturates, we obtain
the additional constraint $0<b<t_{n}$, where $t_{n}$ is the first non-vanishing coefficient of a MacLaurin series for $t_{c}(r)$ [cf. Eq. (14) in~\cite{sergio1}]. From Eqs. (\ref{ard}), (\ref{a1}), to leading order in $r$
we obtain, along the RTG's, \be R\sim r^{\frac{2n}{3}+1}+{\mathcal
O}(r^{p+2-\frac{n}{3}}). \ee This implies $q=2n/3$, and thus the
consistency relation $p=1+q$ reads $p=1+2n/3$, which is formally
the same as that obtained for outgoing radial null geodesics. The
parameters $p$ and $q$ are then uniquely determined from the
initial data, and must obey the constraint $p=1+q\geq
n\,\Rightarrow\,n\leq3$.

The statement in~\cite{sergio1} (second paragraph) that the work of Deshingkar,
Joshi and Dwivedi (DJD) \cite{DJD} shows that ``when one considers {\em timelike} radial
geodesics, the singularity is found to be locally naked and Tipler strong
for an infinite number of non-spacelike geodesics, irrespective of the
initial data'' is partially incorrect: DJD show that only curvature
strength is independent of the initial data, {\em not} visibility.

An additional comment concerns the parameters $a_i$ and $c_i$, introduced
in Eqs. (28)-(34) in~\cite{sergio1}. Since $R''$ is obtained from $R'$ by
differentiation with respect to $r$ along the geodesic, $a_3$ is linearly dependent
on $a_1$: $a_3=qa_1$. Similarly, $c_3=(\alpha p+\beta)c_1$. We note that
the constants $c_i$ are not ``free'', since they must be fixed by
consistency relations involving $R$ and its derivatives. With the
substitutions, the algebraic constraint $C(a_i,c_i)=0$ reads $a_1c_2+2a_2c_1=0$.
That is, for given initial data (whereby $a_1$ and $a_2$ are fixed),
there is only one degree of freedom in the specification of the two $c_i$
parameters (whose ratio is fixed).

As mentioned above, it can be shown that the absence of a radial
null geodesic emerging from a central singularity is sufficient to
guarantee censorship of the singularity, i.e., it rules out the
existence of any causal geodesic emerging from the singularity. To
see this, consider a general spherically symmetric space-time with
line element
\[ ds^2 = -e^{2\mu} dt^2 + e^{2\nu} dr^2 +R^2(r,t)d\Omega^2,\]
where $\mu=\mu(r,t)$, $\nu=\nu(r,t)$. Then the tangent to a causal
geodesic satisfies
\[ -e^{2\mu}{\dot t}^2+e^{2\nu}{\dot r}^2
+\frac{L^2}{R^2}=\epsilon,\] where the overdot represents
differentiation with respect to an affine parameter, $L$ is the
conserved angular momentum, and $\epsilon=0,-1$ for null and
time-like geodesics, respectively. Thus, at any point on such a
geodesic,
\[ e^{2\mu}{\dot t}^2 \geq e^{2\nu}{\dot r}^2,\]
with equality holding {\em only} for radial null geodesics. On the
$t-r$ plane, this reads \be \frac{dr}{dt} \leq
e^{\mu-\nu},\label{ineq1}\ee where we take the positive root for
future pointing outgoing geodesics (we can use coordinate freedom
to guarantee that $t$ increases into the future globally, and
$\partial_rR\geq0$ in a neighbourhood of $R=0$). We can read
(\ref{ineq1}) as \be \frac{dr_{CG}}{dt}<\frac{dr_{RNG}}{dt},
\label{ineq2}\ee where the subscripts represent causal (excluding
radial null) geodesics and outgoing radial null geodesics
respectively. Now suppose that a CG extends back to a central
singularity located on the $t-r$ plane at $r=0$, $t=t_0$. Assume
that the singularity is of the form $t=t_c(r)$ with $t_c(0)=t_0$
and that the regular region of space-time is $t<t_c(r)$. This is
the case for the singularity studied above. Let $p$ be any point
on the CG, to the future of the singularity. Applying the
inequality (\ref{ineq2}) at $p$, we see that the RNG through $p$
crosses CG from below and hence points on this RNG prior to $p$
must lie to the {\em future} of points on CG prior to $p$. Thus,
the RNG, which necessarily lies at $t<t_c(r)$, must extend back to
$r=0$ at time $t=t_0$, and so must emerge from the singularity.
The contrapositive of this result gives the censorship result
mentioned above.

We conclude by emphasising that, whereas
the analysis of~\cite{sergio2}---wherein the general
solution is derived and the singularity is analysed
along radial null directions---is correct, that of~\cite{sergio1} was
incomplete, which led to the incorrect claim that the singularity is always
locally naked along outgoing RTG's,
regardless of the initial data. The assertion, in~\cite{sergio1}, that
the singularity is always Tipler strong along RTG's remains true, and is
independent of the visibility.
We have shown here that the emergence of outgoing
RTG's from the singularity
{\em is} dependent on the initial data, and thus the singularity
is {\em not} always locally naked along RTG's.
In particular, we have shown that initial data that precludes outgoing RNG's also
forbids outgoing RTG's, in any spherically symmetric spacetime.

\section*{Acknowledgments}
BCN acknowledges support from the DCU Albert College Fellowship
scheme. FCM thanks CMAT, U.Minho and FCT (Portugal) for grant
PRAXIS XXI BD/16012/98. SMCVG acknowledges the support of FCT
(Portugal) Grant PRAXIS XXI-BPD-16301-98, and NSF Grants
AST-9731698 and PHY-0099568.

\end{document}